\documentclass{article}
\usepackage{spconf,amsmath,graphicx,hyperref, amssymb}
\usepackage{amsthm}
\newtheorem{proposition}{Proposition}
\usepackage{caption}

\title{Target Detection in Two-Channel Passive Radars with Inter-Receiver Collaboration}
\name{
  Nandan Sriranga,
  Haodong Yang, 
  Venkata Gandikota, \text{and}
  Pramod K. Varshney
}

\address{
  Department of Electrical Engineering and Computer Science, Syracuse University
}
\begin{document}
%
\maketitle
\begin{abstract}
We address the problem of target detection using spatially distributed two-channel passive radars. The transmitter is assumed to be an unknown illuminator-of-opportunity (IO), which transmits a waveform lying in a known low-dimensional subspace (e.g., OFDM). Each receiver transforms its reference and surveillance signals into the IO-subspace, to obtain cross-correlation (CC) measurements. To save bandwidth, receivers collaboratively exchange and linearly combine the CC outputs, after which only a subset of the receivers transmit to a fusion center (FC), over a multiple-access channel (MAC).  The collaboration weights are designed to enhance detection performance. Simulation results are provided to illustrate the performance the proposed framework.       
\end{abstract}
\section{Introduction}
\label{sec:intro}

Passive radar systems exploit signals generated by digital audio broadcasting (DAB) transmitters, digital video broadcasting (DVB) transmitters, and mobile communication transmitters such as WiFi and LTE \cite{berger2010signal}, for target detection and state estimation. They offer practical advantages since dedicated transmitter antennas are not required, but complicates target detection methodology, as the illuminator-of-opportunity (IO) waveform is unknown and can not be designed. This has motivated significant work on detection algorithms and communication-efficient architectures for distributed passive radar systems. To address the issue of the unknown IO waveform, many works assume a known transmitter communication format \cite{polonen2013control, gogineni2017passive, karthik2018improved, horstmann2020two}. Commercial DAB/DVB systems typically employ OFDM or linear digital modulation \cite{berger2010signal, gogineni2017passive}, enabling the IO to be represented as a subspace signal with a known basis \cite{gogineni2017passive, karthik2018improved, mcwhorter2023passive, ramirez2024passive}. This structure has been exploited in detectors such as generalized likelihood-ratio tests (GLRTs) \cite{karthik2018improved, mcwhorter2023passive, ramirez2024passive} and subspace-similarity tests \cite{gogineni2017passive, wang2016canonical, santamaria2017passive}.
In \cite{wang2017signal}, the IO waveform is modeled as an autoregressive process of known order, and the temporal signal correlation of the waveform is estimated and used for target state estimation. In many passive radar systems, receivers are equipped with two channels: (i) a surveillance channel that captures target echoes and (ii) a reference channel that records the direct-path signal from the IO \cite{gogineni2017passive, liu2015performance}. Cross-correlation (CC) between reference and surveillance data is commonly used as a detection statistic \cite{gogineni2017passive, wang2016canonical, santamaria2017passive, wei2021adaptive}. For example, \cite{gogineni2017passive} uses CCs of the principal eigenvectors of reference and surveillance data, while \cite{wang2016canonical, santamaria2017passive} show that the GLRT reduces to canonical cross-correlations estimated from these channels. Several works have also explored the fusion and compression of measurements collected across distributed radar receivers \cite{wei2021adaptive, ma2022compressive}. For example, \cite{wei2021adaptive} sends CCs from each receiver to the FC, where their magnitudes are fused for detection, while \cite{ma2022compressive} compresses receiver measurements before transmission. Our work is primarily motivated by the application of inter-sensor collaboration for estimation, tracking, and detection in distributed sensor networks \cite{kar2013linear, zhang2018optimal, cheng2021joint}, where neighboring sensor nodes exchange measurements locally and then a subset of sensors forward their measurements after linear collaboration to the FC.  

Distributed systems with multiple radar receivers contain a large number of measurements, especially in two-channel passive radar receivers. To reduce communication-bandwidth for receiver-to-FC communication: i) we compute local CC statistics using prior knowledge of the IO waveform subspace at the receivers, and then ii) allow receivers to exchange measurements with neighboring receivers, so that the number of transmitting receivers to the FC can be reduced. The MAC channel to the FC is modeled as a multipl-input single-output channel, and the FC receives a single complex-valued quantity for decision-making. Using the received measurement received at the FC, an energy detector (ED) is used to detect the presence of the target. Based on modeling assumptions in this work, the ED performance depends only on second-order moments, and an appropriate surrogate for detection performance is the ratio of variances, similar to the metric developed in\cite{cheng2021joint}. The framework proposed in this work is aimed at improving uplink communication-efficiency in distributed radar systems.

\section{System Model}
\label{sec:sys_model}
In this work, we consider a passive radar system with $L$ spatially distributed dual-channel receivers. A single IO transmitter which produces a waveform is considered (see Fig. \ref{fig:system_model}). We assume that the direct-path signal has been suppressed in the receiver channel, and clutter-cancellation methods have been applied to the receivers. A total of $N$ snapshots are sampled as measurements in both channels at the $i^{th}$ receiver
{\small
\begin{align}
         \mathcal{H}_1: \hspace{0.1cm} &r_{i}(n) = \beta_{i} \cdot x(n) + \delta_{r,i}(n), \nonumber\\
           &s_{i}(n) = \alpha_i \cdot x(n-\tau_i) \cdot e^{j \boldsymbol{\Omega}_i n} + \delta_{s,i}(n), \nonumber \\
        \mathcal{H}_0: \hspace{0.1cm} &r_{i}(n) = \beta_i \cdot x(n) + \delta_{r,i}(n), \nonumber\\
          & s_{i}(n) =  \delta_{s,i}(n),
    \label{eq:measurements_hyp}  
\end{align}}where $\mathcal{H}_1$ is the target present hypothesis ($\mathcal{H}_0$ denotes no target), $\alpha_i \sim \mathcal{CN}(0, \sigma_{\alpha}^{2})$ is the radar cross-section (RCS) of the target with respect to receiver $i$, $\beta_i \sim \mathcal{CN}(\mu_{\beta}, \sigma_{\beta}^{2})$ is the direct path fading coefficient from the IO to the $i^{th}$ sensor. The direct path signal from the IO to the reference channel usually follows line-of-sight propagation and therefore we assume that the mean $\mu_B \neq 0$. The quantity $\tau_i$ is the propagation delay of the target echo, and $\boldsymbol{\Omega}_i$ is the normalized Doppler frequency at the $i^{th}$ receiver's surveillance channel, where $i=1, \dots, L$. 
\begin{center}
    \centering
    \begin{figure}
    \includegraphics[width = \linewidth, height = 0.3\linewidth, trim=0.2cm 0cm 0cm 0.2cm, clip]{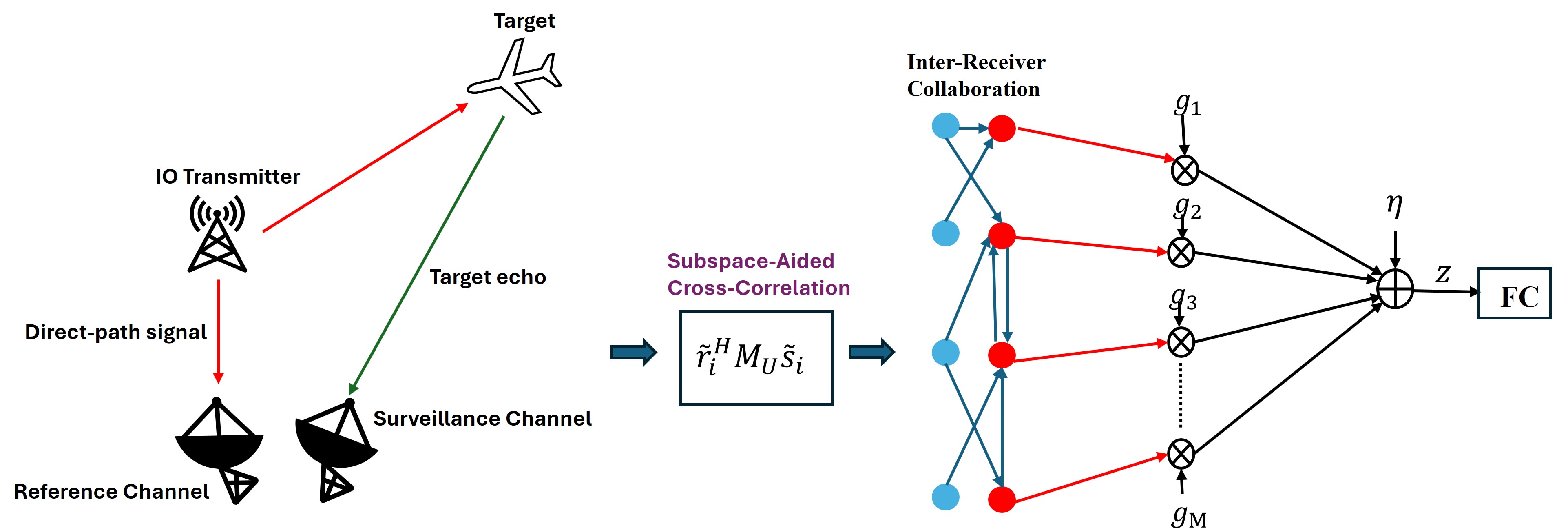}
    \captionsetup{font=scriptsize}
    \caption{Complete system model of a distributed passive-radar system with collaborating receivers. Receivers collaborate with neighboring (spatially proximal) receivers to improve detection performance and communication-efficiency.}
    \label{fig:system_model}
    \end{figure}
\end{center}\vspace{-0.8cm}

Our interest is in detecting the presence of a target, in a specific Delay-Doppler cell (known $\boldsymbol{\Omega}_i$ and $\tau_i$). Therefore, we assume that the Delay-Doppler compensation is enforced perfectly in the surveillance channels. The stacked measurement vectors of the reference and Delay-Doppler compensated surveillance channel at receiver is given as 
{\small
\begin{align}
         \mathcal{H}_1: \hspace{0.1cm} &\mathbf{r}_i = {\beta}_i \cdot \mathbf{x} + \boldsymbol{\delta}_{r,i}, \nonumber\\
           &\mathbf{s}_i = {\alpha}_i \cdot \mathbf{x} + \boldsymbol{\delta}_{s,i}, \nonumber \\
        \mathcal{H}_0: \hspace{0.1cm} &\mathbf{r}_i = \beta_i \cdot \mathbf{x} + \boldsymbol{\delta}_{r,i}, \nonumber\\
          & \mathbf{s}_i = \boldsymbol{\delta}_{s,i},
\label{eq:vec_measurements_hyp}  
\end{align}}where $\mathbf{x} = [x(1) \thinspace x(2) \dots x(N)]^{\mathsf{T}} \in \mathbb{C}^{N}$ is the vector of IO waveform snapshots. The noise vectors are assumed to be independent across different channels and receivers, and are distributed as complex-Gaussian random vectors $\boldsymbol{\delta}_{s,i} \sim\mathcal{CN}(\mathbf{0}, \boldsymbol{\Sigma}_{s})$ and $\boldsymbol{\delta}_{r,i} \sim \mathcal{CN}(\mathbf{0}, \boldsymbol{\Sigma}_{r})$, where the noise covariances $\boldsymbol{\Sigma}_{r}$ and $\boldsymbol{\Sigma}_{s}$ are assumed to be known. The noise in the reference and surveillance channels are assumed to be uncorrelated with each other. We assume that the IO waveform $\mathbf{x}$ maybe represented as \cite{gogineni2017passive, mcwhorter2023passive, ramirez2024passive}
\begin{equation}
    \mathbf{x} = \mathbf{U}_{\text{D}} \boldsymbol{\theta}_{\text{D}},
\end{equation}
where $\boldsymbol{\theta}_{\text{D}} \in \mathbb{C}^{\text{D}}$, and the columns of $\mathbf{U_D}$ span a $D$-dimensional subspace. For example, if the transmitters employ OFDM phase-shift keying, the subspace $\mathbf{U}_{\text{D}}$ is spanned by a subset of the columns of the Fourier basis. Generally, commercial transmitters transmit a diverse range of symbols, due to which the elements of $\boldsymbol{\theta}_{\text{D}}$ are assumed to be random, and the first and second-order statistics are known, i.e., $\mathbb{E}[\boldsymbol{\theta}_{\text{D}}] = \mathbf{0}_{\text{D}}$ and $\text{Cov}(\boldsymbol{\theta}_{\text{D}}) = \boldsymbol{\Sigma}_{\boldsymbol{\theta}}$. Usually, the elements of $\boldsymbol{\theta}_{\text{D}}$ possess phase-shift keying (PSK) or amplitude modulation (AM) alphabets, which are equiprobable and symmetric across the origin.    

In this work, we transform the channel measurements onto the subspace spanned by $\mathbf{U}_{\text{D}}$, thereby exploiting the subspace knowledge at the receivers. Then, CCs are performed after transformation, unlike prior work, where the time-domain measurements are correlated \cite{liu2015performance, wei2021adaptive}. We define the \emph{noise-whitened subspace-transformations} for the reference and surveillance channel measurements as
{\small
\begin{align}
    &\mathbf{T}_{\mathbf{r}} = (\mathbf{U}^{\mathsf{H}}_{\text{D}} \boldsymbol{\Sigma}_{\mathbf{r}}^{-1}\mathbf{U}_{\text{D}})^{-1} \mathbf{U}_{\text{D}}^{\mathsf{H}} \boldsymbol{\Sigma}_{\mathbf{r}}^{-1/2}, \quad \text{and} \nonumber \\
    &\mathbf{T}_{\mathbf{s}} = (\mathbf{U}^{\mathsf{H}}_{\text{D}} \boldsymbol{\Sigma}_{\mathbf{s}}^{-1} \mathbf{U}_{\text{D}})^{-1} \mathbf{U}_{\text{D}}^{\mathsf{H}} \boldsymbol{\Sigma}_{\mathbf{s}}^{-1/2}. 
    \label{eq:transform_subspace}
\end{align}}
The whitened measurements at the $i^{th}$ receiver are $\mathbf{r}_{\text{wh},i} = \boldsymbol{\Sigma}_{\mathbf{r}}^{-1/2}\mathbf{r}_i$ and $\mathbf{s}_{\text{wh},i}=\boldsymbol{\Sigma}_{\mathbf{s}}^{-1/2}\mathbf{s}_i$, at the surveillance and reference channels respectively. Using the transformation matrices defined in (\ref{eq:transform_subspace}), the reference and surveillance channel measurements defined in (\ref{eq:vec_measurements_hyp}) are correlated with each other, after transformation. These are defined as $\tilde{\mathbf{r}}_i = \mathbf{T}_{\mathbf{r}} \mathbf{r}_{\text{wh},i}$ and $\tilde{\mathbf{s}}_i = \mathbf{T}_{\text{s}} \mathbf{s}_{\text{wh},i}$. Using the whitened and subspace-transformed measurements, we define the CCs as
{\small
\begin{equation}
    c_i = \tilde{\mathbf{r}}^{\mathsf{H}}_i (\mathbf{U}^{\mathsf{H}}_{\text{D}} \boldsymbol{\Sigma}_{\mathbf{r}}^{-1/2} \boldsymbol{\Sigma}_{\mathbf{s}}^{-1/2} \mathbf{U}_{\text{D}})\tilde{\mathbf{s}}_i.
    \label{eq:cross_correl}
\end{equation}
}
The expression in (\ref{eq:cross_correl}) computes the CC using the whitened and subspace-transformed coordinates. The CC between $\tilde{\mathbf{r}_i}$ and $\tilde{\mathbf{s}_i}$, involves the matrix $\mathbf{M}_{\mathbf{U}} = \mathbf{U}^{\mathsf{H}}_{\text{D}} \boldsymbol{\Sigma}_{\mathbf{r}}^{-1/2} \boldsymbol{\Sigma}_{\mathbf{s}}^{-1/2} \mathbf{U}_{\text{D}}$, so that the signal component of the CC in (\ref{eq:cross_correl}) has the same strength as the time domain CC output, due to the subspace structure. In contrast, the noise components which are outside the signal subspace are eliminated after subspace transformation. 

Upon computing the cross-correlation terms based on (\ref{eq:cross_correl}) at each receiver, a subset $\mathcal{S}_{M}$ of $M$ receivers out of the $L$ receivers procure measurements from their neighbors and linearly combine the CC terms $c_i$ as
{\small
\begin{equation}
    y_i = \sum_{k \in \mathcal{N}_i} w_{ik} c_k + w_{ii} c_{i} + \epsilon_i, 
\end{equation}}where $w_{ik}$ are the complex weights used by the $k^{th}$ neighbor to forward their measurements to receiver $i$. The collaboration between receivers induces noise $\epsilon_i \sim \mathcal{CN}(0, \sigma_{\epsilon}^{2})$ at each of the $M$ transmitting sensors. The post-collaboration vector is denoted as 
\begin{equation}
  \mathbf{y} = \mathbf{W} \mathbf{c} + \boldsymbol{\epsilon},
  \label{eq:post_collab_vec}
\end{equation}
where the matrix $\mathbf{W} \in \mathbb{C}^{\text{M} \times \text{L}}$ is governed by
$\mathbf{W} \odot (\mathbf{1}_{\text{M}}\mathbf{1}^{\mathsf{T}}_{\text{L}}-\mathbf{A}) = \mathbf{0}_{\text{ML}}$. The operator $\odot$ is the element-wise Hadamard product, $\mathbf{A}_{ik} = 1$ if the receiver $k$ transmits the CC term $c_k$ to the receiver with index $i$, otherwise $\mathbf{A}_{ik} = 0$. Also, since every sensor in the subset $\mathcal{S}_{M}$ takes part in the collaboration process, $A_{ii} = 1 \hspace{0.1cm} \forall i \in \mathcal{S}_{M}$. Essentially, $\mathbf{A}$ is a matrix denoting the communication graph between neighboring receivers. 

\section{Target Detection at Fusion Center}
\label{sec:targ_det}
The uplink communication channel used by receivers to report to the FC is modeled as a noisy linear multiple-access channel (MAC), to account for realistic communication between remote receivers and the FC. After collaboration, we assume that the receivers transmit their measurements over the MAC channel to the FC. The measurements received at the FC are
\begin{equation}
    z = \mathbf{g}^{\mathsf{H}} \mathbf{y} + \eta,
    \label{eq:mac_scalar_FC}
\end{equation}
where $\mathbf{g} \in \mathbb{C}^{\text{M}}$ is the known complex channel-gain for the MAC channel. The MAC channel noise is complex-Gaussian with $\eta \sim \mathcal{CN}(0, \sigma_{\eta}^{2})$. It is difficult to assume a specific distribution for the measurement $z$. Therefore, we resort to an energy detector of the form 
{\small
\begin{equation}
    \vert z \vert^{2} \hspace{0.15cm} \mathop{\gtrless}\limits_{\mathcal{H}_0}^{\mathcal{H}_1} \hspace{0.15cm} \tau,
    \label{eq:energy_detector}
\end{equation}}where $\vert\cdot \vert$ is the magnitude of a complex quantity, and $\tau$ is the decision threshold. The detection performance can be characterized by the first and second-order moments of the measurement $z$ at the FC. If $D$ is reasonably large, the IO waveform $\mathbf{x}$ can be approximated as an $N$-dimensional complex-Gaussian vector. Using this, it is easy to see that the means $\mathbb{E}[c_i| \mathcal{H}_1] = \mathbb{E}[c_i| \mathcal{H}_0] = 0$ and the variances are stated in the Proposition~\ref{prop:sec_momnent} below. 
\begin{proposition}\label{prop:sec_momnent}
The second-order moments $\sigma_{c,1}^{2} = \mathrm{Var}(c_i|H_1)$, $\sigma_{c,0}^{2} = \mathrm{Var}(c_i|H_0)$ are stated as
{\small
\begin{align}
\vspace{-0.5cm}
\sigma_{c,0}^{2}
&= (\mu^{2}_{\beta}+\sigma_{\beta}^2) \,
\mathrm{tr}\!\left(\mathbf{M}_{\mathbf{U}}\mathbf{C}_s\mathbf{M}_{\mathbf{U}}^{\mathsf{H}}\boldsymbol{\Sigma}_{\theta}\right)
   + \mathrm{tr}\!\left(\mathbf{M}_{\mathbf{U}}\mathbf{C}_s\mathbf{M}_{\mathbf{U}}^{\mathsf{H}}\mathbf{C}_r\right), \label{eq:var_ci_H0}\\
   \vspace{-0.5cm}
\sigma_{c,1}^{2}
&= \mathrm{Var}(c_i \mid \mathcal{H}_0)
   + \sigma_{\alpha}^2 \,     \mathrm{tr}\!\left(\mathbf{M}_{\mathbf{U}}^{\mathsf{H}}\mathbf{C}_r\mathbf{M}_{\mathbf{U}}\boldsymbol{\Sigma}_{\theta}\right) \notag \\[6pt]
&\quad
   + \sigma_{\alpha}^2 (\mu^{2}_{\beta}+\sigma_{\beta}^2)  \,
     \Big( \mathrm{tr}\!\left(\mathbf{M}_{\mathbf{U}}\boldsymbol{\Sigma}_{\theta}\mathbf{M}_{\mathbf{U}}^{\mathsf{H}}\boldsymbol{\Sigma}_{\theta}\right)
     + \big|\mathrm{tr}(\mathbf{M}_{\mathbf{U}}\boldsymbol{\Sigma}_{\theta})\big|^2 \Big),
   \label{eq:var_ci_H1}
\end{align}
}
with the definitions $\mathbf{C}_r = \left( \mathbf{U}_{\text{D}}^{\mathsf{H}} 
\boldsymbol{\Sigma}_{\mathbf{r}}^{-1} 
\mathbf{U}_{\text{D}} \right)^{-1}$ and
$\mathbf{C}_s = \left( \mathbf{U}_{\text{D}}^{\mathsf{H}} 
\boldsymbol{\Sigma}_{\mathbf{s}}^{-1} 
\mathbf{U}_{\text{D}} \right)^{-1}$. It can also be shown that 
\(
\mathrm{Cov}(c_i, c_j|\mathcal{H}_{\ell}) = 0, \hspace{0.1cm} \text{for } \thinspace l \in\{0,1\} \thinspace \text{and} \thinspace i \neq j.
\)
To derive the expression in (\ref{eq:var_ci_H0}), we first consider $\mathrm{Cov}(c_i, c_j|\mathcal{H}_{\ell})$ for all $i \neq j$. 

We use the fact that $c_i \;=\; \mathbf{r}_i^{H}\,\mathbf{A}\,\mathbf{s}_i,$ and define $\mathbf{A} \;\triangleq\; 
\boldsymbol{\Sigma}_{r}^{-1/2}\,\mathbf{U}_{D}\,\mathbf{C}_{r}\,\mathbf{U}_{D}^{\mathsf H}\,\boldsymbol{\Sigma}_{r}^{-1/2}\,
\boldsymbol{\Sigma}_{s}^{-1/2}\,\mathbf{U}_{D}\,\mathbf{C}_{s}\,\mathbf{U}_{D}^{\mathsf H}\,\boldsymbol{\Sigma}_{s}^{-1/2}$. Using the property of total covariance, 
\begin{align}
\operatorname{Cov}(c_i,c_j\mid \mathcal{H}_{\ell})
&= \mathbb{E}\!\big[\operatorname{Cov}(c_i,c_j\mid \mathbf{x},\mathcal{H}_{\ell})\big] \nonumber \\
 &+ \operatorname{Cov}\!\big(\mathbb{E}[c_i\mid \mathbf{x},\mathcal{H}_{\ell}],\,\mathbb{E}[c_j\mid \mathbf{x},\mathcal{H}_{\ell}]\big).
 \label{eq:law_total_cov}
\end{align}We first compute the covariance term under $\mathcal{H}_0$, where the target echo is absent in the surveillance channel, with $c_i = (\beta_i \mathbf{x} + \boldsymbol{\delta}_{r,i})^{H} \mathbf{A} \boldsymbol{\delta}_{s,i}$ and $c_j = (\beta_j \mathbf{x} + \boldsymbol{\delta}_{r,j})^{H} \mathbf{A} \boldsymbol{\delta}_{s,j}$.  
\begin{align*}
\mathbb{E}[c_i c_j^{*}\mid \mathbf{x},\mathcal{H}_0]
&= \mathbb{E}\!\Big[(\beta_i^{*}\mathbf{x}^{H}\mathbf{A}\,\boldsymbol{\delta}_{s,i} 
     + \boldsymbol{\delta}_{r,i}^{H}\mathbf{A}\,\boldsymbol{\delta}_{s,i}) \\
&\quad \cdot (\boldsymbol{\delta}_{s,j}^{H}\mathbf{A}^{H}\beta_j \mathbf{x} 
     + \boldsymbol{\delta}_{s,j}^{H}\mathbf{A}^{H}\boldsymbol{\delta}_{r,j}) \mid \mathbf{x}\Big],\\
\mathrm{Cov}(c_i,c_j\mid \mathbf{x},\mathcal{H}_0)
&= \mathbb{E}[c_i c_j^{*}\mid \mathbf{x},\mathcal{H}_0] \\
&\quad - \mathbb{E}[c_i\mid \mathbf{x},\mathcal{H}_0]\,
         \mathbb{E}[c_j^{*}\mid \mathbf{x},\mathcal{H}_0]. 
\end{align*}
We compute the conditional mean of $c_i$'s as
\begin{align}
\mathbb{E}[c_i \mid \mathbf{x},\mathcal{H}_0]
&= \mathbb{E}\!\Big[(\beta_i \mathbf{x}+\boldsymbol{\delta}_{r,i})^{H}\mathbf{A}\,\boldsymbol{\delta}_{s,i}\,\Big|\ \mathbf{x}\Big] \nonumber \\[2pt]
&= \mathbb{E}[\beta_i^{*}]\,\mathbf{x}^{H}\mathbf{A}\,\underbrace{\mathbb{E}[\boldsymbol{\delta}_{s,i}]}_{=\,0}
 \;+\; \underbrace{\mathbb{E}\!\big[\boldsymbol{\delta}_{r,i}^{H}\mathbf{A}\,\boldsymbol{\delta}_{s,i}\big]}_{=\ \mathrm{tr}\!\left(\mathbf{A}\,\mathbb{E}[\boldsymbol{\delta}_{s,i}\boldsymbol{\delta}_{r,i}^{H}]\right)=\,0} \nonumber\\[2pt]
&= 0,
\label{eq:condn_mean_H_0}
\end{align}
where the expectation term inside the trace is $\mathbf{0}$ due to the assumption of independence across reference and surveilance channel measurements in the same receiver. Then, we expand the conditional expectation term 
\begin{align}
&\mathbb{E}[c_i c_j^{*}\mid \mathbf{x},\mathcal{H}_0] \nonumber\\
&= \mathbb{E}\!\Big[(\beta_i \mathbf{x}+\boldsymbol{\delta}_{r,i})^{H}\mathbf{A}\,\boldsymbol{\delta}_{s,i}\,
                     \boldsymbol{\delta}_{s,j}^{H}\mathbf{A}^{H}(\beta_j \mathbf{x}+\boldsymbol{\delta}_{r,j})\Big] \nonumber\\[2pt]
&= \underbrace{\mathbb{E}\!\big[\beta_i^{*}\beta_j\big]\,
   \mathbf{x}^{H}\mathbf{A}\,\mathbb{E}\!\big[\boldsymbol{\delta}_{s,i}\boldsymbol{\delta}_{s,j}^{H}\big]\,
   \mathbf{A}^{H}\mathbf{x}}_{(1)} \nonumber\\[2pt]
&\quad + \underbrace{\mathbf{x}^{H}\mathbf{A}\,
   \mathbb{E}\!\big[\beta_i^{*}\boldsymbol{\delta}_{s,i}\boldsymbol{\delta}_{s,j}^{H}\mathbf{A}^{H}\boldsymbol{\delta}_{r,j}\big]}_{(2)} \nonumber\\[2pt]
&\quad + \underbrace{\mathbb{E}\!\big[\boldsymbol{\delta}_{r,i}^{H}\mathbf{A}\boldsymbol{\delta}_{s,i}\boldsymbol{\delta}_{s,j}^{H}\big]\,
   \mathbf{A}^{H}\beta_j \mathbf{x}}_{(3)} \nonumber\\[2pt]
&\quad + \underbrace{\mathbb{E}\!\big[\boldsymbol{\delta}_{r,i}^{H}\mathbf{A}\boldsymbol{\delta}_{s,i}\boldsymbol{\delta}_{s,j}^{H}\mathbf{A}^{H}\boldsymbol{\delta}_{r,j}\big]}_{(4)}.
\label{eq:condn_crrl_H_0}
\end{align}
Each of the four terms may be simplified as 
\begin{align*}
&\mathbb{E}[\beta_i^{*}\beta_j]\,
  \mathbf{x}^{H}\mathbf{A}\,\mathbb{E}[\boldsymbol{\delta}_{s,i}\boldsymbol{\delta}_{s,j}^{H}]\,\mathbf{A}^{H}\mathbf{x} \\
&= \mathbb{E}\!\Big[
   \mathbb{E}[\beta_i^{*}\beta_j \mid \boldsymbol{\delta}_{s,i},\boldsymbol{\delta}_{s,j}]\,
   \mathbf{x}^{H}\mathbf{A}\,\boldsymbol{\delta}_{s,i}\boldsymbol{\delta}_{s,j}^{H}\,\mathbf{A}^{H}\mathbf{x}\Big] \\
&= \mathbb{E}[\beta_i^{*}]\mathbb{E}[\beta_j]\,
   \mathbf{x}^{H}\mathbf{A}\,\mathbb{E}[\boldsymbol{\delta}_{s,i}\boldsymbol{\delta}_{s,j}^{H}]\,\mathbf{A}^{H}\mathbf{x} \;=\; 0,
\end{align*}
where $\mathbb{E}[\boldsymbol{\delta}_{s,i}\boldsymbol{\delta}_{s,j}^{H}] = \mathbf{0}$, as the surveillance channel noise across receivers $\boldsymbol{\delta}_{s,i}$ and $\boldsymbol{\delta}_{s,j}$, are uncorrelated. Term (2) is simplified as
\begin{align*}
&\mathbf{x}^{H}\mathbf{A}\,
  \mathbb{E}[\beta_i^{*}\boldsymbol{\delta}_{s,i}\boldsymbol{\delta}_{s,j}^{H}\mathbf{A}^{H}\boldsymbol{\delta}_{r,j}] \\
&= \mathbf{x}^{H}\mathbf{A}\,
   \mathbb{E}\!\Big[
   \mathbb{E}[\beta_i^{*}\boldsymbol{\delta}_{s,i}\boldsymbol{\delta}_{s,j}^{H}\mathbf{A}^{H}\boldsymbol{\delta}_{r,j}
   \mid \boldsymbol{\delta}_{s,i},\boldsymbol{\delta}_{s,j}]\Big] \\
&= \mathbf{x}^{H}\mathbf{A}\,
   \mathbb{E}\!\Big[\beta_i^{*}\boldsymbol{\delta}_{s,i}\boldsymbol{\delta}_{s,j}^{H}\mathbf{A}^{H}\,
   \underbrace{\mathbb{E}[\boldsymbol{\delta}_{r,j}\mid \boldsymbol{\delta}_{s,i},\boldsymbol{\delta}_{s,j}]}_{=\;\mathbf{0}}\Big]
   \;=\; 0.
\end{align*}
Simplifying term (3) yields
\begin{align*}
&\mathbb{E}[\boldsymbol{\delta}_{r,i}^{H}\mathbf{A}\boldsymbol{\delta}_{s,i}\boldsymbol{\delta}_{s,j}^{H}]\,
  \mathbf{A}^{H}\beta_j\mathbf{x} \\
&= \mathbb{E}\!\Big[
   \mathbb{E}[\boldsymbol{\delta}_{r,i}^{H}\mathbf{A}\boldsymbol{\delta}_{s,i}\boldsymbol{\delta}_{s,j}^{H}
   \mid \boldsymbol{\delta}_{r,i},\boldsymbol{\delta}_{s,i}]\Big]\,
   \mathbf{A}^{H}\beta_j\mathbf{x} \\
&= \mathbb{E}\!\Big[\boldsymbol{\delta}_{r,i}^{H}\mathbf{A}\boldsymbol{\delta}_{s,i}\,
   \underbrace{\mathbb{E}[\boldsymbol{\delta}_{s,j}^{H}\mid \boldsymbol{\delta}_{r,i},\boldsymbol{\delta}_{s,i}]}_{=\;\mathbf{0}}\Big]\,
   \mathbf{A}^{H}\beta_j\mathbf{x} \;=\; 0.
\end{align*}
Similarly, term (4) may also be simplified as 
\begin{align*}
&\mathbb{E}[\boldsymbol{\delta}_{r,i}^{H}\mathbf{A}\boldsymbol{\delta}_{s,i}\boldsymbol{\delta}_{s,j}^{H}\mathbf{A}^{H}\boldsymbol{\delta}_{r,j}] \\
&= \mathbb{E}\!\Big[
   \mathbb{E}[\boldsymbol{\delta}_{r,i}^{H}\mathbf{A}\boldsymbol{\delta}_{s,i}\boldsymbol{\delta}_{s,j}^{H}\mathbf{A}^{H}\boldsymbol{\delta}_{r,j}
   \mid \boldsymbol{\delta}_{r,i},\boldsymbol{\delta}_{s,i},\boldsymbol{\delta}_{s,j}]\Big] \\
&= \mathbb{E}\!\Big[\boldsymbol{\delta}_{r,i}^{H}\mathbf{A}\boldsymbol{\delta}_{s,i}\boldsymbol{\delta}_{s,j}^{H}\mathbf{A}^{H}\,
   \underbrace{\mathbb{E}[\boldsymbol{\delta}_{r,j}\mid \boldsymbol{\delta}_{r,i},\boldsymbol{\delta}_{s,i},\boldsymbol{\delta}_{s,j}]}_{=\;\mathbf{0}}\Big]
   \;=\; 0.
\end{align*}
Since both expectation terms (\ref{eq:condn_mean_H_0}) and (\ref{eq:condn_crrl_H_0}) are individually equal to $0$, the quantity $\operatorname{Cov}(c_i,c_j\mid \mathcal{H}_0) = 0$. We now consider the covariance term under $\mathcal{H}_1$, where the target echo is present in the surveillance channel, with 
$c_i = (\beta_i \mathbf{x} + \boldsymbol{\delta}_{r,i})^{H} \mathbf{A} (\alpha_i \mathbf{x} + \boldsymbol{\delta}_{s,i})$ and 
$c_j = (\beta_j \mathbf{x} + \boldsymbol{\delta}_{r,j})^{H} \mathbf{A} (\alpha_j \mathbf{x} + \boldsymbol{\delta}_{s,j})$.  
\begin{align*}
&\mathbb{E}[c_i c_j^{*}\mid \mathbf{x},\mathcal{H}_1] \\
&= \mathbb{E}\!\Big[(\beta_i^{*}\mathbf{x}^{H}\mathbf{A}(\alpha_i\mathbf{x}+\boldsymbol{\delta}_{s,i})
     + \boldsymbol{\delta}_{r,i}^{H}\mathbf{A}(\alpha_i\mathbf{x}+\boldsymbol{\delta}_{s,i})) \\
&\quad \cdot ((\alpha_j\mathbf{x}+\boldsymbol{\delta}_{s,j})^{H}\mathbf{A}^{H}\beta_j \mathbf{x} 
     + (\alpha_j\mathbf{x}+\boldsymbol{\delta}_{s,j})^{H}\mathbf{A}^{H}\boldsymbol{\delta}_{r,j}) \,\Big|\mathbf{x}\Big],\\
&\mathrm{Cov}(c_i,c_j\mid \mathbf{x},\mathcal{H}_1) \\
&= \mathbb{E}[c_i c_j^{*}\mid \mathbf{x},\mathcal{H}_1] 
 - \mathbb{E}[c_i\mid \mathbf{x},\mathcal{H}_1]\,
   \mathbb{E}[c_j^{*}\mid \mathbf{x},\mathcal{H}_1]. 
\end{align*}
We compute the conditional mean of $c_i$ as
\begin{align}
\mathbb{E}[c_i \mid \mathbf{x},\mathcal{H}_1]
&= \mathbb{E}\!\Big[(\beta_i \mathbf{x}+\boldsymbol{\delta}_{r,i})^{H}\mathbf{A}(\alpha_i \mathbf{x}+\boldsymbol{\delta}_{s,i})\;\Big|\;\mathbf{x}\Big] \nonumber\\[2pt]
&= \mathbb{E}[\beta_i^{*}\alpha_i]\,\mathbf{x}^{H}\mathbf{A}\mathbf{x} 
 + \mathbb{E}[\beta_i^{*}]\,\mathbf{x}^{H}\mathbf{A}\,\mathbb{E}[\boldsymbol{\delta}_{s,i}] \\
 &+ \mathbb{E}[\alpha_i]\,\mathbb{E}[\boldsymbol{\delta}_{r,i}^{H}]\mathbf{A}\mathbf{x}
 + \mathbb{E}[\boldsymbol{\delta}_{r,i}^{H}\mathbf{A}\boldsymbol{\delta}_{s,i}] \nonumber\\[2pt]
&= 0,
\label{eq:condn_mean_H_1}
\end{align}
where all expectations disappear due to zero-mean and independence assumptions. Then, we expand the conditional expectation term 
\begin{align}
&\mathbb{E}[c_i c_j^{*}\mid \mathbf{x},\mathcal{H}_1] \nonumber\\
=&\mathbb{E}\!\Big[(\beta_i \mathbf{x}+\boldsymbol{\delta}_{r,i})^{H}\mathbf{A}(\alpha_i\mathbf{x}+\boldsymbol{\delta}_{s,i})\,
(\alpha_j\mathbf{x}+\boldsymbol{\delta}_{s,j})^{H}\mathbf{A}^{H} (\beta_j \mathbf{x}+\boldsymbol{\delta}_{r,j})\Big] \nonumber\\[2pt]
=& \underbrace{\mathbb{E}[\beta_i^{*}\alpha_i\alpha_j^{*}\beta_j]\,
   \mathbf{x}^{H}\mathbf{A}\mathbf{x}\,\mathbf{x}^{H}\mathbf{A}^{H}\mathbf{x}}_{(1)}  + \underbrace{\mathbf{x}^{H}\mathbf{A}\,
   \mathbb{E}[\beta_i^{*}\alpha_i\mathbf{x}\,\boldsymbol{\delta}_{s,j}^{H}\mathbf{A}^{H}\boldsymbol{\delta}_{r,j}]}_{(2)} \nonumber\\[2pt]
&\quad + \underbrace{\mathbb{E}[\boldsymbol{\delta}_{r,i}^{H}\mathbf{A}\alpha_i\mathbf{x}\,\alpha_j^{*}\mathbf{x}^{H}\mathbf{A}^{H}\boldsymbol{\delta}_{r,j}]}_{(3)} \nonumber\\[2pt]
&\quad + \underbrace{\mathbb{E}[\boldsymbol{\delta}_{r,i}^{H}\mathbf{A}\boldsymbol{\delta}_{s,i}\,\boldsymbol{\delta}_{s,j}^{H}\mathbf{A}^{H}\boldsymbol{\delta}_{r,j}]}_{(4)}.
\label{eq:condn_crrl_H_1}
\end{align}
Each of the four terms may be simplified as 
\begin{align*}
&\mathbb{E}[\beta_i^{*}\alpha_i\alpha_j^{*}\beta_j]\,
\mathbf{x}^{H}\mathbf{A}\mathbf{x}\,\mathbf{x}^{H}\mathbf{A}^{H}\mathbf{x} \\
&= \mathbb{E}[\beta_i^{*}]\mathbb{E}[\beta_j]\mathbb{E}[\alpha_i]\mathbb{E}[\alpha_j^{*}]
\,\mathbf{x}^{H}\mathbf{A}\mathbf{x}\,\mathbf{x}^{H}\mathbf{A}^{H}\mathbf{x} = 0,
\end{align*}
since $\alpha_i,\alpha_j$ are zero-mean and independent. Term (2) simplifies as
\begin{align*}
&\mathbf{x}^{H}\mathbf{A}\,
   \mathbb{E}[\beta_i^{*}\alpha_i\mathbf{x}\,\boldsymbol{\delta}_{s,j}^{H}\mathbf{A}^{H}\boldsymbol{\delta}_{r,j}] \\
&= \mathbf{x}^{H}\mathbf{A}\,
   \mathbb{E}\!\Big[\beta_i^{*}\alpha_i\mathbf{x}\,\boldsymbol{\delta}_{s,j}^{H}\mathbf{A}^{H}\,
   \underbrace{\mathbb{E}[\boldsymbol{\delta}_{r,j}\mid \alpha_i,\boldsymbol{\delta}_{s,j}]}_{=\;\mathbf{0}}\Big] = 0.
\end{align*}
Term (3) becomes
\begin{align*}
\mathbb{E}[\boldsymbol{\delta}_{r,i}^{H}\mathbf{A}\alpha_i\mathbf{x}\,\alpha_j^{*}\mathbf{x}^{H}\mathbf{A}^{H}\boldsymbol{\delta}_{r,j}]
&= \mathbb{E}\!\Big[\boldsymbol{\delta}_{r,i}^{H}\mathbf{A}\mathbf{x}\,\mathbf{x}^{H}\mathbf{A}^{H}\boldsymbol{\delta}_{r,j}\,
   \underbrace{\mathbb{E}[\alpha_i\alpha_j^{*}]}_{=0}\Big] = 0.
\end{align*}
Finally, term (4) is simplified as
\begin{align*}
&\mathbb{E}[\boldsymbol{\delta}_{r,i}^{H}\mathbf{A}\boldsymbol{\delta}_{s,i}\,\boldsymbol{\delta}_{s,j}^{H}\mathbf{A}^{H}\boldsymbol{\delta}_{r,j}] \\
&= \mathbb{E}\!\Big[\boldsymbol{\delta}_{r,i}^{H}\mathbf{A}\boldsymbol{\delta}_{s,i}\,\boldsymbol{\delta}_{s,j}^{H}\mathbf{A}^{H}\,
   \underbrace{\mathbb{E}[\boldsymbol{\delta}_{r,j}\mid \boldsymbol{\delta}_{r,i},\boldsymbol{\delta}_{s,i},\boldsymbol{\delta}_{s,j}]}_{=\;\mathbf{0}}\Big] = 0.
\end{align*}
Since both expectation terms (\ref{eq:condn_mean_H_1}) and (\ref{eq:condn_crrl_H_1}) are individually equal to $0$, the quantity 
$\operatorname{Cov}(c_i,c_j\mid \mathcal{H}_1) = 0$ follows immediately from the law of total covariance in (\ref{eq:law_total_cov}). 
Then, we proceed to compute $\operatorname{Var}(c_i \mid \mathcal{H}_{\ell})$, for $\ell \in \{0,1\}$. We use the complex-Gaussian moment-generating functions in order to arrive at the expressions. We consider the random vectors
{\small
\begin{align*}
\mathbf{z}\sim\mathcal{CN}(\mathbf{0},\boldsymbol{\Sigma}),\ 
\mathbf{u}\sim\mathcal{CN}(\mathbf{0},\boldsymbol{\Sigma}_u),\ 
\mathbf{v}\sim\mathcal{CN}(\mathbf{0},\boldsymbol{\Sigma}_v),\ 
\end{align*}
}where $\mathbf{u}$ and $\mathbf{v}$ are mutually independent random vectors. For the quadratic form $Q=\mathbf{z}^{H}\mathbf{M}\mathbf{z}$, the moment generating function (MGF) is
{\small
\begin{align*}
M_Q(t)=\mathbb{E}[e^{tQ}]
= \det\!\big(\mathbf{I}-t\,\boldsymbol{\Sigma}\mathbf{M}\big)^{-1}.
\end{align*}
}
We begin with a proper complex Gaussian vector $\mathbf{z}\sim\mathcal{CN}(\mathbf{0},\boldsymbol{\Sigma})$. The moment generating function (MGF) can be derived by Gaussian integration

{\small
\begin{align*}
\mathbb{E}[e^{tQ}]
&=\frac{1}{\pi^n\det(\boldsymbol{\Sigma})}\int_{\mathbb{C}^n}\exp\!\Big(-\mathbf{z}^{H}\boldsymbol{\Sigma}^{-1}\mathbf{z}+t\,\mathbf{z}^{H}\mathbf{M}\mathbf{z}\Big)\,d\mathbf{z}\\
&=\frac{\det(\boldsymbol{\Sigma}^{-1})}{\det(\boldsymbol{\Sigma}^{-1}-t\mathbf{M})}
=\det(\mathbf{I}-t\,\boldsymbol{\Sigma}\mathbf{M})^{-1}.
\end{align*}
}
We use the cumulant generating function
$K_Q(t)=\log M_Q(t)=-\log\det(\mathbf{I}-t\,\boldsymbol{\Sigma}\mathbf{M})$, to simplify the derivation. From the definition of the cumulant generating function, $K_Q'(0)=\mathbb{E}[Q]$ and $K_Q''(0)=\mathrm{Var}(Q)$. Upon differentiation,
{\small
\begin{align*}
K_Q'(t)&=\mathrm{tr}\!\Big((\mathbf{I}-t\,\boldsymbol{\Sigma}\mathbf{M})^{-1}\boldsymbol{\Sigma}\mathbf{M}\Big),\\
K_Q''(t)&=\mathrm{tr}\!\Big((\mathbf{I}-t\,\boldsymbol{\Sigma}\mathbf{M})^{-1}\boldsymbol{\Sigma}\mathbf{M}(\mathbf{I}-t\,\boldsymbol{\Sigma}\mathbf{M})^{-1}\boldsymbol{\Sigma}\mathbf{M}\Big).
\end{align*}
}
Evaluating this at $t=0$ gives the well-known expressions for the mean and variance of the quadratic form:

{\small
\begin{align*}
\mathbb{E}[Q]=\mathrm{tr}(\mathbf{M}\boldsymbol{\Sigma}),\qquad
\mathrm{Var}(Q)=\mathrm{tr}(\mathbf{M}\boldsymbol{\Sigma}\mathbf{M}\boldsymbol{\Sigma}).
\end{align*}
}

Next, to compute the second moment of the modulus, we introduce the joint MGF of $Q$ and $Q^{*}$,
$M(s,t)=\mathbb{E}[e^{sQ+tQ^{*}}]$, where $Q^{*}=\mathbf{z}^{H}\mathbf{M}^{H}\mathbf{z}$. 
This evaluates to
{\small
\begin{align*}
M(s,t)=\det(\mathbf{I}-\boldsymbol{\Sigma}(s\mathbf{M}+t\mathbf{M}^{H}))^{-1}.
\end{align*}
}We consider $K(s,t) = \operatorname{log}(M(s,t))$, and differentiate with respect both to $s$ and $t$. This yields
{\small
\begin{align*}
&K^{'}_s(0,0)=\mathrm{tr}(\boldsymbol{\Sigma}\mathbf{M}),\quad
K^{'}_t(0,0)=\mathrm{tr}(\boldsymbol{\Sigma}\mathbf{M}^{H}),\\
&K^{''}_{st}(0,0)=\mathrm{tr}(\boldsymbol{\Sigma}\mathbf{M}^{H}\boldsymbol{\Sigma}\mathbf{M}).
\end{align*}
}
Using the chain rule for MGFs, $M_{st}(0,0)=K^{''}_{st}(0,0)+K^{'}_s(0,0)K^{''}_t(0,0)$, and therefore
{\small
\begin{align*}
\mathbb{E}[|Q|^{2}]
=\mathrm{tr}(\mathbf{M}\boldsymbol{\Sigma}\mathbf{M}^{H}\boldsymbol{\Sigma})+\big|\mathrm{tr}(\mathbf{M}\boldsymbol{\Sigma})\big|^{2}.
\end{align*}
}Then, we consider the bilinear form $B=\mathbf{u}^{H}\mathbf{M}\mathbf{v}$, 
where $\mathbf{u}\sim\mathcal{CN}(\mathbf{0},\boldsymbol{\Sigma}_{u})$ 
and $\mathbf{v}\sim\mathcal{CN}(\mathbf{0},\boldsymbol{\Sigma}_{v})$ are independent. 
We stack them into $\mathbf{w}=\begin{bmatrix}\mathbf{u}\\\mathbf{v}\end{bmatrix}$ 
with block covariance $\boldsymbol{\Sigma}_{w}=\mathrm{diag}(\boldsymbol{\Sigma}_{u},\boldsymbol{\Sigma}_{v})$, 
and define 
$\mathbf{K}=\begin{bmatrix}0&\mathbf{M}\\0&0\end{bmatrix}$ 
so that $B=\mathbf{w}^{H}\mathbf{K}\mathbf{w}$. 
The joint MGF $M(s,t)=\det(\mathbf{I}-s\,\boldsymbol{\Sigma}_{w}\mathbf{K}-t\,\boldsymbol{\Sigma}_{w}\mathbf{K}^{H})^{-1}$, which simplifies to
{\small
\begin{align*}
&K_s(0,0)=0,\quad K_t(0,0)=0,\\
&K_{st}(0,0)=\mathrm{tr}(\boldsymbol{\Sigma}_{v}\mathbf{M}^{H}\boldsymbol{\Sigma}_{u}\mathbf{M}).
\end{align*}
}Since $\mathbb{E}[B]=0$, the quantity
{\small
\begin{align*}
\mathbb{E}[|B|^{2}]
=\mathrm{tr}(\mathbf{M}\boldsymbol{\Sigma}_{v}\mathbf{M}^{H}\boldsymbol{\Sigma}_{u}).
\end{align*}
}We use the variances of the quadratic form and the bilinear form to derive the expressions stated in Prop. \ref{prop:sec_momnent}.

The variance under $\mathcal{H}_0$ is first considered. The reference and surveillance channel measurements respectively are $\mathbf{r}_i=\beta_i\mathbf{x}+\boldsymbol{\delta}_{r,i}$, $\mathbf{s}_i=\boldsymbol{\delta}_{s,i}$, where $\mathbf{x}\sim\mathcal{CN}(\mathbf{0},\boldsymbol{\Sigma}_x)$ is independent of all other quantities (noise, target RCS and direct-path fading). Since $c_i=\mathbf{r}_i^{H}\mathbf{A}\mathbf{s}_i$, under $\mathcal{H}_0$, it assumes the form 
{\small
\begin{align*}
c_i&=(\beta_i\mathbf{x}+\boldsymbol{\delta}_{r,i})^{H}\mathbf{A}\,\boldsymbol{\delta}_{s,i}
=\underbrace{\beta_i^{*}\,\mathbf{x}^{H}\mathbf{A}\boldsymbol{\delta}_{s,i}}_{T_1}
+\underbrace{\boldsymbol{\delta}_{r,i}^{H}\mathbf{A}\boldsymbol{\delta}_{s,i}}_{T_2}. 
\end{align*}
}
We then compute the covariance of the terms $T_1$ and $T_2$ as 
{\small
\begin{align*}
\mathbb{E}[T_1 T_2^{*}]
&=\mathbb{E}\!\Big[\beta_i^{*}\,\mathbf{x}^{H}\mathbf{A}\boldsymbol{\delta}_{s,i}\,
                     \boldsymbol{\delta}_{s,i}^{H}\mathbf{A}^{H}\boldsymbol{\delta}_{r,i}\Big]\\
&=\mathbb{E}\!\Big[\,
   \mathbb{E}\!\big[\beta_i^{*}\,\mathbf{x}^{H}\mathbf{A}\boldsymbol{\delta}_{s,i}\,
                     \boldsymbol{\delta}_{s,i}^{H}\mathbf{A}^{H}\boldsymbol{\delta}_{r,i}
                     \,\big|\,\beta_i,\mathbf{x},\boldsymbol{\delta}_{s,i}\big]\Big]\\
&=\mathbb{E}\!\Big[\beta_i^{*}\,\mathbf{x}^{H}\mathbf{A}\boldsymbol{\delta}_{s,i}\,
                    \boldsymbol{\delta}_{s,i}^{H}\mathbf{A}^{H}\,
                    \underbrace{\mathbb{E}[\boldsymbol{\delta}_{r,i}\mid \beta_i,\mathbf{x},\boldsymbol{\delta}_{s,i}]}_{=\;\mathbf{0}}\Big]\\
&=0.
\end{align*}
}
The expression for the variance is 
{\small
\begin{align*}
\mathrm{Var}(c_i\mid\mathcal{H}_0)
&=\mathbb{E}[|c_i|^2\mid\mathcal{H}_0] \\\
&=\mathbb{E}[|T_1|^2]+\mathbb{E}[|T_2|^2] + \mathbb{E}[T_1 T^{*}_2] \\
&= \mathbb{E}[|T_1|^2]+\mathbb{E}[|T_2|^2] 
\end{align*}
}since $\mathbb{E}[T_1] = \mathbb{E}[T_2] = 0$ and $\mathbb{E}[T_1 T^{*}_{2}] = 0$. Next, we consider the terms
\begin{align*}
    \mathbb{E}[|T_1|^2]
&=\mathbb{E}[|\beta_i|^2]\;\mathbb{E}[|\mathbf{x}^{H}\mathbf{A}\boldsymbol{\delta}_{s,i}|^2]\\
&\overset{(a)}{=}(\mu_\beta^2+\sigma_\beta^2)\;\mathrm{tr}\!\left(\mathbf{A}\boldsymbol{\Sigma}_s\mathbf{A}^{H}\boldsymbol{\Sigma}_{\mathbf{x}}\right),\\[6pt]
\mathbb{E}[|T_2|^2]
&=\mathbb{E}\!\Big[\big(\boldsymbol{\delta}_{r,i}^{H}\mathbf{A}\boldsymbol{\delta}_{s,i}\big)
                   \big(\boldsymbol{\delta}_{r,i}^{H}\mathbf{A}\boldsymbol{\delta}_{s,i}\big)^{*}\Big]\\
&=\mathbb{E}\!\Big[\boldsymbol{\delta}_{r,i}^{H}\mathbf{A}\boldsymbol{\delta}_{s,i}\,
                   \boldsymbol{\delta}_{s,i}^{H}\mathbf{A}^{H}\boldsymbol{\delta}_{r,i}\Big]\\
&=\mathbb{E}\!\Big[\mathrm{tr}\!\big(\mathbf{A}\boldsymbol{\delta}_{s,i}\boldsymbol{\delta}_{s,i}^{H}\mathbf{A}^{H}\boldsymbol{\delta}_{r,i}\boldsymbol{\delta}_{r,i}^{H}\big)\Big]\\
&=\mathrm{tr}\!\Big(\mathbf{A}\,\mathbb{E}[\boldsymbol{\delta}_{s,i}\boldsymbol{\delta}_{s,i}^{H}]\,\mathbf{A}^{H}\,
                     \mathbb{E}[\boldsymbol{\delta}_{r,i}\boldsymbol{\delta}_{r,i}^{H}]\Big)\\
&\overset{(b)}{=}\mathrm{tr}\!\left(\mathbf{A}\boldsymbol{\Sigma}_s\mathbf{A}^{H}\boldsymbol{\Sigma}_r\right),
\end{align*} 
where $(a)$ is due to the variance of the bilinear form and $(b)$ is due to the law of iterated expectations. Then, the variance can be written as
\begin{align}
    \sigma_{c,0}^{2}
=(\mu_\beta^2+\sigma_\beta^2)\,\mathrm{tr}\!\left(\mathbf{A}\boldsymbol{\Sigma}_s\mathbf{A}^{H}\boldsymbol{\Sigma}_x\right)
+\mathrm{tr}\!\left(\mathbf{A}\boldsymbol{\Sigma}_s\mathbf{A}^{H}\boldsymbol{\Sigma}_r\right).
\label{eq:var_c_i_H_0}
\end{align}This is simplified using the definitions of $\mathbf{A}$ and $\boldsymbol{\Sigma}_{\mathbf{x}}$ to
{\small
\begin{align*}
\mathbf{A} &= \boldsymbol{\Sigma}_{r}^{-1/2}\,\mathbf{U}_{D}\,\mathbf{C}_{r}\,\mathbf{U}_{D}^{H}\,\boldsymbol{\Sigma}_{r}^{-1/2}\,
             \boldsymbol{\Sigma}_{s}^{-1/2}\,\mathbf{U}_{D}\,\mathbf{C}_{s}\,\mathbf{U}_{D}^{H}\,\boldsymbol{\Sigma}_{s}^{-1/2}, \\[4pt]
\mathbf{C}_{r} &= \big(\mathbf{U}_{D}^{H}\,\boldsymbol{\Sigma}_{r}^{-1}\,\mathbf{U}_{D}\big)^{-1}, \qquad
\mathbf{C}_{s} = \big(\mathbf{U}_{D}^{H}\,\boldsymbol{\Sigma}_{s}^{-1}\,\mathbf{U}_{D}\big)^{-1}, \\[4pt]
\boldsymbol{\Sigma}_{\mathbf{x}} &= \mathbf{U}_{D}\,\boldsymbol{\Sigma}_{\boldsymbol{\theta}}\,\mathbf{U}_{D}^{H}, \qquad
\mathbf{M}_{\mathbf{U}} = \mathbf{U}_{D}^{H}\,\boldsymbol{\Sigma}_{r}^{-1/2}\,\boldsymbol{\Sigma}_{s}^{-1/2}\,\mathbf{U}_{D}.
\end{align*}
}Using these definitions, the trace terms in  (\ref{eq:var_c_i_H_0}) are simplified to
\begin{align*}
\mathrm{tr}\!\left(\mathbf{A}\,\boldsymbol{\Sigma}_{s}\,\mathbf{A}^{H}\,\boldsymbol{\Sigma}_{\mathbf{x}}\right)
&= \mathrm{tr}\!\left(\mathbf{M}_{\mathbf{U}}\,\mathbf{C}_{s}\,\mathbf{M}_{\mathbf{U}}^{H}\,\boldsymbol{\Sigma}_{\boldsymbol{\theta}}\right), \\[6pt]
\mathrm{tr}\!\left(\mathbf{A}\,\boldsymbol{\Sigma}_{s}\,\mathbf{A}^{H}\,\boldsymbol{\Sigma}_{r}\right)
&= \mathrm{tr}\!\left(\mathbf{M}_{\mathbf{U}}\,\mathbf{C}_{s}\,\mathbf{M}_{\mathbf{U}}^{H}\,\mathbf{C}_{r}\right), \\[6pt]
\mathrm{tr}\!\left(\mathbf{A}\,\boldsymbol{\Sigma}_{\mathbf{x}}\,\mathbf{A}^{H}\,\boldsymbol{\Sigma}_{\mathbf{x}}\right)
&= \mathrm{tr}\!\left(\mathbf{M}_{\mathbf{U}}\,\boldsymbol{\Sigma}_{\boldsymbol{\theta}}\,\mathbf{M}_{\mathbf{U}}^{H}\,\boldsymbol{\Sigma}_{\boldsymbol{\theta}}\right), \\[6pt]
\big|\mathrm{tr}(\mathbf{A}\,\boldsymbol{\Sigma}_{\mathbf{x}})\big|^{2}
&= \big|\mathrm{tr}(\mathbf{M}_{\mathbf{U}}\,\boldsymbol{\Sigma}_{\boldsymbol{\theta}})\big|^{2}.
\end{align*}
Using similar arguments we consider the variance under $\mathcal{H}_1$ of the $c_i$ terms. The reference and surveillance channel measurements are $\mathbf{r}_i=\beta_i\mathbf{x}+\boldsymbol{\delta}_{r,i}$, $\mathbf{s}_i=\alpha_i\mathbf{x}+\boldsymbol{\delta}_{s,i}$, where $\mathbf{x}\sim\mathcal{CN}(\mathbf{0},\boldsymbol{\Sigma}_{\mathbf{x}})$ is independent of all other quantities. Using $c_i=\mathbf{r}_i^{H}\mathbf{A}\mathbf{s}_i$, under $\mathcal{H}_1$ it assumes the form 
{\small
\begin{align*}
c_i&=(\beta_i\mathbf{x}+\boldsymbol{\delta}_{r,i})^{H}\mathbf{A}\,(\alpha_i\mathbf{x}+\boldsymbol{\delta}_{s,i}) \\
&=\underbrace{\beta_i^{*}\alpha_i\,\mathbf{x}^{H}\mathbf{A}\mathbf{x}}_{U_1}
+\underbrace{\beta_i^{*}\,\mathbf{x}^{H}\mathbf{A}\boldsymbol{\delta}_{s,i}}_{U_2} \\
&+\underbrace{\boldsymbol{\delta}_{r,i}^{H}\mathbf{A}\alpha_i\mathbf{x}}_{U_3}
+\underbrace{\boldsymbol{\delta}_{r,i}^{H}\mathbf{A}\boldsymbol{\delta}_{s,i}}_{U_4}.
\end{align*}
}We first verify that the covariances of any pair $U_p$, $U_q$ for $p \neq q$ are 0. We illustrate as an example, using the law of iterated expectations,
{\small
\begin{align*}
\mathbb{E}[U_2 U_4^{*}]
&=\mathbb{E}\!\Big[\beta_i^{*}\,\mathbf{x}^{H}\mathbf{A}\boldsymbol{\delta}_{s,i}\,
                     \boldsymbol{\delta}_{s,i}^{H}\mathbf{A}^{H}\boldsymbol{\delta}_{r,i}\Big]\\
&=\mathbb{E}\!\Big[\,
   \mathbb{E}\!\big[\beta_i^{*}\,\mathbf{x}^{H}\mathbf{A}\boldsymbol{\delta}_{s,i}\,
                     \boldsymbol{\delta}_{s,i}^{H}\mathbf{A}^{H}\boldsymbol{\delta}_{r,i}
                     \,\big|\,\beta_i,\mathbf{x},\boldsymbol{\delta}_{s,i}\big]\Big]\\
&=\mathbb{E}\!\Big[\beta_i^{*}\,\mathbf{x}^{H}\mathbf{A}\boldsymbol{\delta}_{s,i}\,
                    \boldsymbol{\delta}_{s,i}^{H}\mathbf{A}^{H}\,
                    \underbrace{\mathbb{E}[\boldsymbol{\delta}_{r,i}\mid \beta_i,\mathbf{x},\boldsymbol{\delta}_{s,i}]}_{=\;\mathbf{0}}\Big]=0.
\end{align*}
}Similar arguments may be used to show that $\mathbb{E}[U_p U_q^{*}]=0$ for all $p\neq q$. Hence,
{\small
\begin{align*}
\mathrm{Var}(c_i\mid\mathcal{H}_1)
&=\mathbb{E}[|c_i|^2\mid\mathcal{H}_1]
=\sum_{k=1}^{4}\mathbb{E}[|U_k|^2].
\end{align*}
}

We expand each of the four terms to complete the derivation. The terms $U_2$ and $U_4$ are the same as the terms $T_1$ and $T_2$ respectively, used for the calculations of the variance under $\mathcal{H}_0$
{\small
\begin{align*}
\mathbb{E}[|U_2|^2]
&=\mathbb{E}[|\beta_i|^2]\;\mathbb{E}[|\mathbf{x}^{H}\mathbf{A}\boldsymbol{\delta}_{s,i}|^2]
=(\mu_\beta^2+\sigma_\beta^2)\;\mathrm{tr}\!\left(\mathbf{A}\boldsymbol{\Sigma}_s\mathbf{A}^{H}\boldsymbol{\Sigma}_{\mathbf{x}}\right),\\[6pt]
\mathbb{E}[|U_4|^2]
&=\mathbb{E}\!\Big[\big(\boldsymbol{\delta}_{r,i}^{H}\mathbf{A}\boldsymbol{\delta}_{s,i}\big)
                   \big(\boldsymbol{\delta}_{r,i}^{H}\mathbf{A}\boldsymbol{\delta}_{s,i}\big)^{*}\Big]\\
&=\mathbb{E}\!\Big[\boldsymbol{\delta}_{r,i}^{H}\mathbf{A}\boldsymbol{\delta}_{s,i}\,
                   \boldsymbol{\delta}_{s,i}^{H}\mathbf{A}^{H}\boldsymbol{\delta}_{r,i}\Big]\\
&=\mathbb{E}\!\Big[\mathrm{tr}\!\big(\mathbf{A}\boldsymbol{\delta}_{s,i}\boldsymbol{\delta}_{s,i}^{H}\mathbf{A}^{H}\boldsymbol{\delta}_{r,i}\boldsymbol{\delta}_{r,i}^{H}\big)\Big]
=\mathrm{tr}\!\left(\mathbf{A}\boldsymbol{\Sigma}_s\mathbf{A}^{H}\boldsymbol{\Sigma}_r\right).
\end{align*}
}The term $U_3$ is a bilinear form scaled by $\alpha_i$, yielding
{\small
\begin{align*}
\mathbb{E}[|U_3|^2]
&=\mathbb{E}\!\Big[\big(\boldsymbol{\delta}_{r,i}^{H}\mathbf{A}\alpha_i\mathbf{x}\big)
                    \big(\boldsymbol{\delta}_{r,i}^{H}\mathbf{A}\alpha_i\mathbf{x}\big)^{*}\Big]\\
&=\mathbb{E}[|\alpha_i|^2]\;\mathbb{E}\!\Big[\boldsymbol{\delta}_{r,i}^{H}\mathbf{A}\mathbf{x}\,
                    \mathbf{x}^{H}\mathbf{A}^{H}\boldsymbol{\delta}_{r,i}\Big]\\
&=\sigma_{\alpha}^{2}\;\mathbb{E}\!\Big[\mathrm{tr}\!\big(\mathbf{A}\mathbf{x}\mathbf{x}^{H}\mathbf{A}^{H}\boldsymbol{\delta}_{r,i}\boldsymbol{\delta}_{r,i}^{H}\big)\Big]
=\sigma_{\alpha}^{2}\;\mathrm{tr}\!\left(\mathbf{A}^{H}\boldsymbol{\Sigma}_r\mathbf{A}\boldsymbol{\Sigma}_{\mathbf{x}}\right).
\end{align*}
}Lastly, $U_1$ is a quadratic form in $\mathbf{x}$ (scaled by $\beta_i^{*}\alpha_i$), so using the complex Gaussian quadratic-form moment,
{\small
\begin{align*}
\mathbb{E}[|U_1|^2]
&=\mathbb{E}[|\beta_i|^2]\;\mathbb{E}[|\alpha_i|^2]\;
  \mathbb{E}\!\Big[\,\big|\mathbf{x}^{H}\mathbf{A}\mathbf{x}\big|^{2}\,\Big]\\
&=(\mu_\beta^2+\sigma_\beta^2)\,\sigma_{\alpha}^{2}\,
  \Big(\mathrm{tr}\!\left(\mathbf{A}\boldsymbol{\Sigma}_{\mathbf{x}}\mathbf{A}^{H}\boldsymbol{\Sigma}_{\mathbf{x}}\right)
      +\big|\mathrm{tr}(\mathbf{A}\boldsymbol{\Sigma}_{\mathbf{x}})\big|^{2}\Big).
\end{align*}
}Summing all the four terms, the variance under $\mathcal{H}_1$ is
{\small
\begin{align}
\sigma_{c,1}^{2}
&=\mathrm{Var}(c_i\mid\mathcal{H}_1) \nonumber\\
&=(\mu_\beta^2+\sigma_\beta^2)\,\mathrm{tr}\!\left(\mathbf{A}\boldsymbol{\Sigma}_s\mathbf{A}^{H}\boldsymbol{\Sigma}_{\mathbf{x}}\right)
+\mathrm{tr}\!\left(\mathbf{A}\boldsymbol{\Sigma}_s\mathbf{A}^{H}\boldsymbol{\Sigma}_r\right) \nonumber\\
&\quad + \sigma_{\alpha}^{2}\,\mathrm{tr}\!\left(\mathbf{A}^{H}\boldsymbol{\Sigma}_r\mathbf{A}\boldsymbol{\Sigma}_{\mathbf{x}}\right)
+(\mu_\beta^2+\sigma_\beta^2)\,\sigma_{\alpha}^{2}\,
 \Big(\mathrm{tr}\!\left(\mathbf{A}\boldsymbol{\Sigma}_{\mathbf{x}}\mathbf{A}^{H}\boldsymbol{\Sigma}_{\mathbf{x}}\right) \nonumber\\
&\quad+\big|\mathrm{tr}(\mathbf{A}\boldsymbol{\Sigma}_{\mathbf{x}})\big|^{2}\Big).
\label{eq:var_c_i_H_1}
\end{align}
}Using $\boldsymbol{\Sigma}_{\mathbf{x}}=\mathbf{U}_{D}\boldsymbol{\Sigma}_{\boldsymbol{\theta}}\mathbf{U}_{D}^{H}$, 
$\mathbf{M}_{\mathbf{U}}=\mathbf{U}_{D}^{H}\boldsymbol{\Sigma}_{r}^{-1/2}\boldsymbol{\Sigma}_{s}^{-1/2}\mathbf{U}_{D}$ and the definition of $\mathbf{A}$, the two trace terms $U_3$ and $U_1$ in \eqref{eq:var_c_i_H_1}, reduce to
{\small
\begin{align*}
\mathrm{tr}\!\left(\mathbf{A}^{H}\boldsymbol{\Sigma}_r\mathbf{A}\boldsymbol{\Sigma}_{\mathbf{x}}\right)
&= \mathrm{tr}\!\left(\mathbf{M}_{\mathbf{U}}^{H}\mathbf{C}_{r}\mathbf{M}_{\mathbf{U}}\boldsymbol{\Sigma}_{\boldsymbol{\theta}}\right),\\[6pt]
\mathrm{tr}\!\left(\mathbf{A}\boldsymbol{\Sigma}_{\mathbf{x}}\mathbf{A}^{H}\boldsymbol{\Sigma}_{\mathbf{x}}\right)
&= \mathrm{tr}\!\left(\mathbf{M}_{\mathbf{U}}\,\boldsymbol{\Sigma}_{\boldsymbol{\theta}}\,\mathbf{M}_{\mathbf{U}}^{H}\,\boldsymbol{\Sigma}_{\boldsymbol{\theta}}\right) \\
\big|\mathrm{tr}(\mathbf{A}\boldsymbol{\Sigma}_{\mathbf{x}})\big|^{2}
&= \big|\mathrm{tr}(\mathbf{M}_{\mathbf{U}}\,\boldsymbol{\Sigma}_{\boldsymbol{\theta}})\big|^{2},
\end{align*}
}
while the first two trace terms are exactly those already simplified under $\mathcal{H}_0$.

\end{proposition}\vspace{-0.2cm}
Since the target RCS at each receiver $\alpha_i$, and all noise terms are zero-mean, and the processing is linear,
the CC scalars $c_i$ are zero-mean under both hypotheses. Hence,
$\mathbb{E}[\mathbf{y}\mid\mathcal{H}_\ell]=\mathbf{0}$ and
$\mathbb{E}[z\mid\mathcal{H}_\ell]=0$ for $\ell\in\{0,1\}$. Consequently, using the expressions in (\ref{eq:post_collab_vec}) and (\ref{eq:mac_scalar_FC}), 
\(
\mathrm{Var}\!\left(z \mid \mathcal{H}_\ell\right)
= \mathbf{g}^{\mathsf{H}} \mathbf{W}\, (\sigma_{c,l}^{2} \mathbf{I}_L) \mathbf{W}^{\mathsf{H}} \mathbf{g}
+ \sigma_{\epsilon}^{2}\,\|\mathbf{g}\|_{2}^{2}
+ \sigma_{\eta}^{2},
\)
where $\ell \in \{0,1\}$ is the hypothesis. 

\begin{figure*}
\centering
\begin{minipage}[htb]{0.45\textwidth}
    \centering
    \includegraphics[width = 0.9\linewidth, height = 0.45\linewidth, trim=0cm 7cm 0cm 7cm, clip]{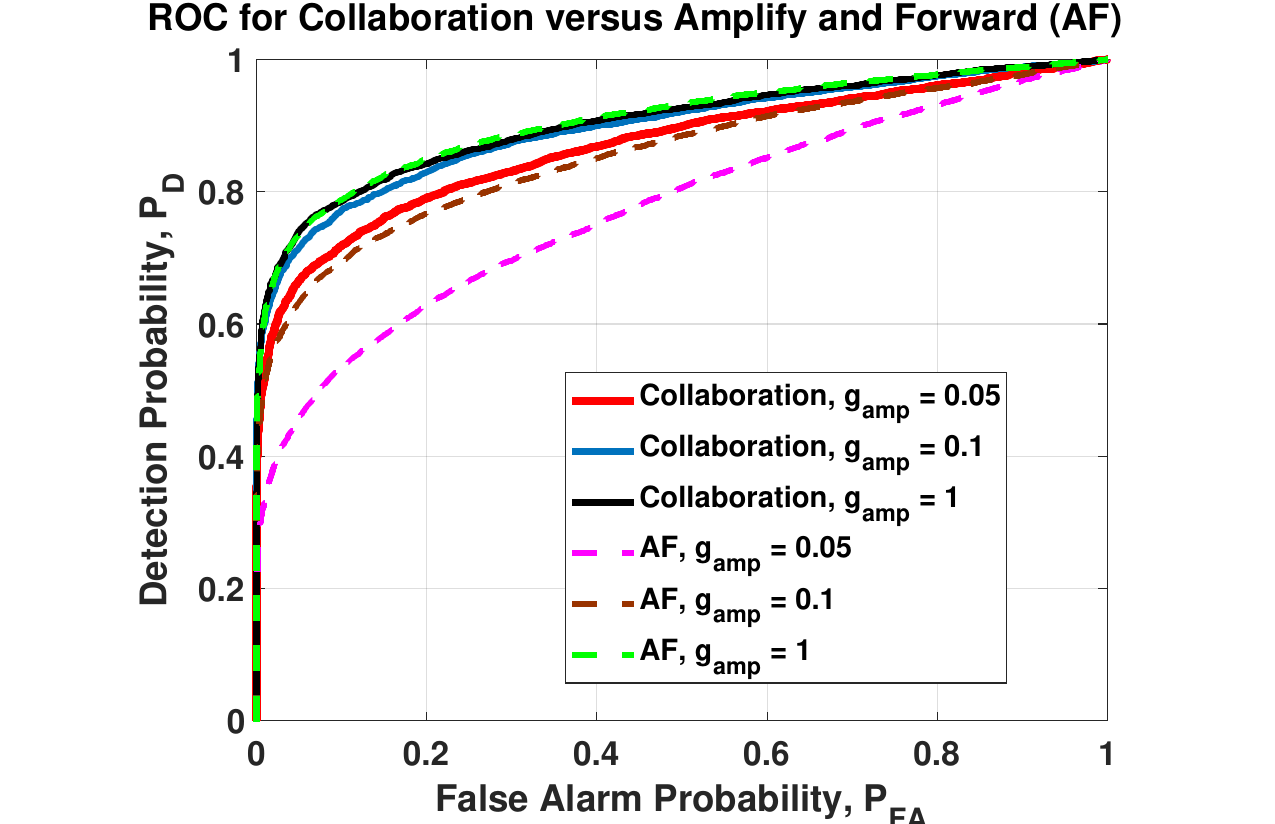}
    \captionsetup{font=scriptsize}
    \caption{The number of transmitting receivers for the plots with collaboration is $M = 5$. The collaboration noise variance is $\sigma_{\epsilon}^{2} = 1$ and the MAC channel noise variance is $\sigma_{\eta}^{2} = 1$. The target RCS variance $\sigma_{\alpha}^{2} = 1$. The collaboration power budget $P_W = 1$ for all plots.}
    \label{fig:collab_AF}
    \end{minipage}\hfill
\begin{minipage}[htb]{0.45\textwidth}
    \centering
    \includegraphics[width = 0.785\linewidth, height = 0.52\linewidth, trim=0cm 5cm 0cm 5.2cm, clip]{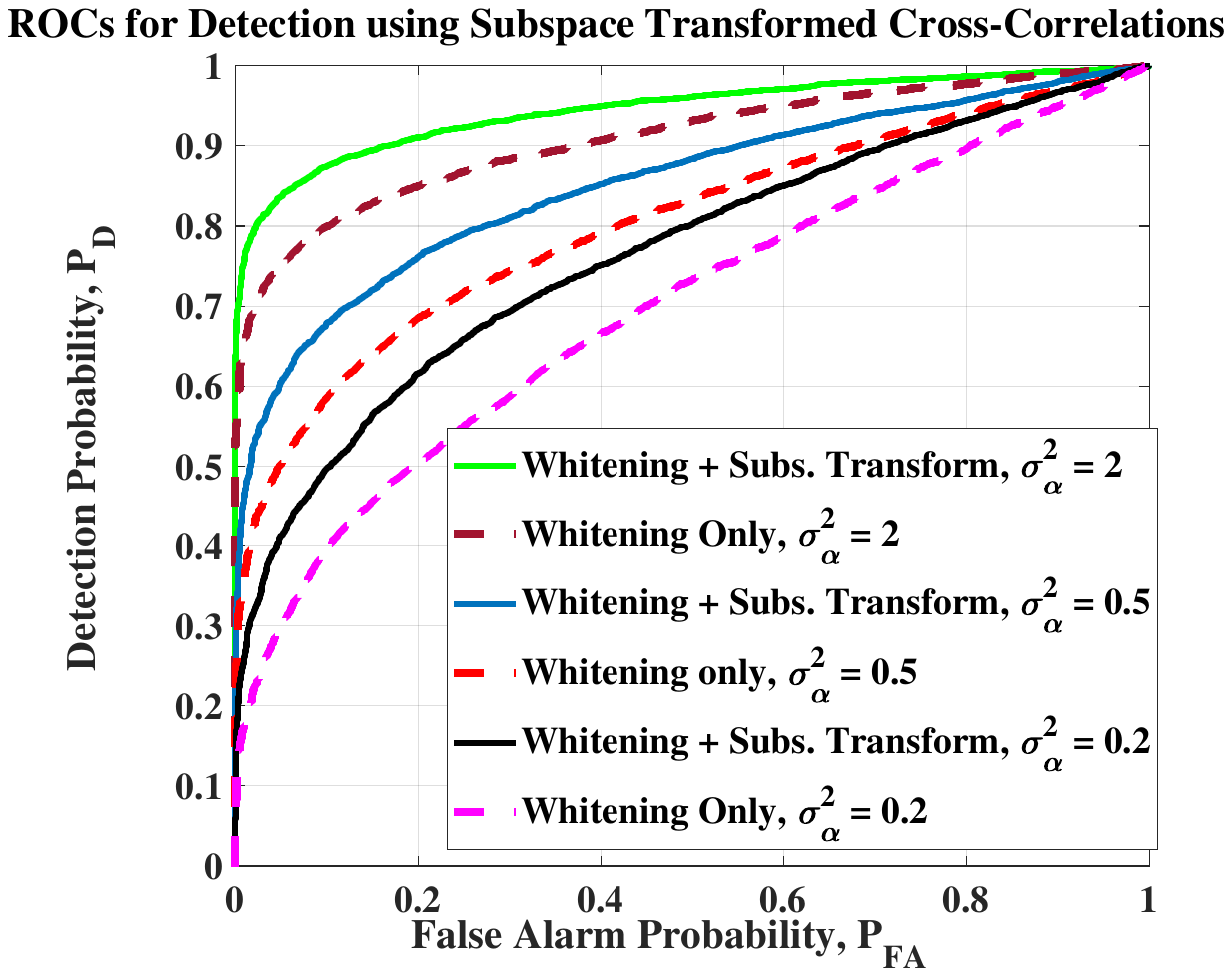}
    \captionsetup{font=scriptsize}
    \caption{Detection performance when receiver measurements use subspace transformations in addition to whitening. We assume that $\sigma_{\text{out}}^{2} = 1$, $\sigma_{\text{in}}^{2} = 2$, $\mu_{\beta} = 1$ and $\sigma_{\beta}^{2}= 1$ for all plots. The number of transmitting sensors for all plots is $M = 5$. The collaboration power budget $P_W = 1$ for all plots.}
    \label{fig:subspace_correl}
\end{minipage}
\end{figure*}\vspace{-0.2cm}

\vspace{-0.2cm}
\section{Inter-Receiver Collaboration Design}
\label{sec:collab_design}
Despite the assumption that IO signal $\mathbf{x}$ is complex-Gaussian, each of the CC terms in equation~\ref{eq:cross_correl} is a quadratic form of a complex-Gaussian vector. Therefore, it is difficult to derive the exact detection performance in terms of the type-I and type-II errors. Since the measurement $z$ is zero-mean under both hypotheses, the ratio of its variances may be adopted as a surrogate metric for detection performance. We define
{\small
\begin{equation}
R(\mathbf{W}) \;\triangleq\; \frac{\mathrm{Var}(z\mid\mathcal{H}_1)}{\mathrm{Var}(z\mid\mathcal{H}_0)},
\end{equation}}which depends only on second-order moments. We remark that the variance ratio $R(\mathbf{W})$ is similar in form to the \emph{generalized-deflection coefficient} defined in \cite{cheng2021joint}. For a fixed \text{collaboration budget} $P_W$, the collaboration matrix $\mathbf{W}$ is designed using 
{\small
\begin{align}
    \underset{\mathbf{W}}{\text{max}} \quad R(\mathbf{W}) \;\triangleq\; & \frac{\mathbf{g}^{\mathsf{H}} \mathbf{W}\,(\sigma_{c,1}^{2} \mathbf{I}_L)\mathbf{W}^{\mathsf{H}} \mathbf{g}
+ \sigma_{\epsilon}^{2}\,\|\mathbf{g}\|_{2}^{2}
+ \sigma_{\eta}^{2}}{\mathbf{g}^{\mathsf{H}} \mathbf{W}\,(\sigma_{c,0}^{2} \mathbf{I}_L)\mathbf{W}^{\mathsf{H}} \mathbf{g}
+ \sigma_{\epsilon}^{2}\,\|\mathbf{g}\|_{2}^{2}
+ \sigma_{\eta}^{2}} \nonumber\\
    \text{subject to} \quad &\text{tr}(\mathbf{W}\mathbf{W}^{H}) = P_{\mathbf{W}}, \hspace{0.1cm} \text{and} \nonumber\\
  &\mathbf{W} \odot (\mathbf{1}_{\text{M}} \mathbf{1}^{\mathsf{T}}_{\text{L}}-\mathbf{A}) = \mathbf{0}_{\text{ML}}.
  \label{eq:var_ratio_maxim}
\end{align}}

To enforce the collaboration topology constraint, we first vectorize $\mathbf{W}$ by dropping all elements from it that correspond to $0$ in $\mathbf{A}$. The aim is to represent this vector-matrix product as $\mathbf{g}^{H} \mathbf{W} = \mathbf{w}^{\mathsf{H}} \mathbf{G}$,
where $\mathbf{G} \in \mathbb{C}^{n_{\text{W}} \times L}$ is a matrix, with $n_{\text{W}}$ being the number of non-zero elements in $\mathbf{W}$. We vectorize the collaboration matrices $\mathbf{W}$ by traversing across the columns (starting from column 1). We define a mapping from the indices of the vector $\mathbf{w}$ to the original indices in $\mathbf{W}$ such that $[\mathbf{w}]_{l} \to [\mathbf{W}]_{i_{l}j_{l}}, \hspace{0.1cm} \text{only if} \thinspace [\mathbf{A}]_{i_l j_l} = 1$
with $l=1, \dots, n_{W}$. Then, the elements of $\mathbf{G}$ are defined according to
{\small
\begin{equation}
    [\mathbf{G}]_{lm} = 
\begin{cases}
    [\mathbf{g}]_{i_l}, & m= j_l  \\
0 & \text{otherwise}.
\end{cases} 
\label{eq:G_def}
\end{equation}}This allows us to discard the non-zero entries in the matrices $\mathbf{W}$ and vectorize it into a vector $\mathbf{w} \in \mathbb{C}^{n_W}$. The elements of the matrix $\mathbf{G}$ are based on (\ref{eq:G_def}). The optimization problem in (\ref{eq:var_ratio_maxim}) is reformulated as
{\small
\begin{equation}
R(\mathbf{w})=\frac{\sigma_{c,1}^{2}\,\mathbf{w}^{\mathsf H}\boldsymbol{\Gamma}\mathbf{w}+\omega}
{\sigma_{c,0}^{2}\,\mathbf{w}^{\mathsf H}\boldsymbol{\Gamma}\mathbf{w}+\omega},
\quad
\text{s.t. } \mathbf{w}^{\mathsf H}\mathbf{w}=P_{\mathbf W},
\label{eq:quadratic_ratio}
\end{equation}} 
where $\boldsymbol{\Gamma} \;\triangleq\; \mathbf{G}\mathbf{G}^{\mathsf H}$ and, $\omega \;\triangleq\; \sigma_{\epsilon}^{2}\|\mathbf{g}\|_2^{2}+\sigma_{\eta}^{2}$.
Multiplying the constant $\omega$ with $\mathbf{w}^{\mathsf H}\mathbf{w}/P_{\mathbf W}=1$ allows us to express the objective in (\ref{eq:quadratic_ratio}) as
{\small
\begin{equation}
R(\mathbf{w})
=\frac{\mathbf{w}^{\mathsf H}\Big(\sigma_{c,1}^{2}\boldsymbol{\Gamma}+\tfrac{\omega}{P_{\mathbf W}}\mathbf{I}_{n_W}\Big)\mathbf{w}}
{\mathbf{w}^{\mathsf H}\Big(\sigma_{c,0}^{2}\boldsymbol{\Gamma}+\tfrac{\omega}{P_{\mathbf W}}\mathbf{I}_{n_W}\Big)\mathbf{w}},
\label{eq:rayleigh_quotient}
\end{equation}}
where the following matrices are defined as
$\boldsymbol{\Omega}_{1}\;\triangleq\;\sigma_{c,1}^{2}\boldsymbol{\Gamma}+\tfrac{\omega}{P_{\mathbf W}}\mathbf{I}_{n_W}$, $\boldsymbol{\Omega}_{0}\;\triangleq\;\sigma_{c,0}^{2}\boldsymbol{\Gamma}+\tfrac{\omega}{P_{\mathbf W}}\mathbf{I}_{n_W}$. Since $\omega>0$, both $\boldsymbol{\Omega}_1$ and $\boldsymbol{\Omega}_0$ are \emph{Hermitian positive-definite}. Using the normalization $\tilde{\mathbf{w}} = \frac{\mathbf{w}}{\sqrt{P_W}}$, the optimization problem simplifies to a \emph{generalized Rayleigh quotient}
{\small
\begin{equation}
\max_{\Vert \Vert\mathbf{\tilde{w}} \Vert \Vert_2 = 1}\ \ \frac{\mathbf{\tilde{w}}^{\mathsf H}\boldsymbol{\Omega}_{1}\mathbf{\tilde{w}}}{\mathbf{\tilde{w}}^{\mathsf H}\boldsymbol{\Omega}_{0}\mathbf{\tilde{w}}}
\quad\triangleq\quad
\boldsymbol{\Omega}_{1}\mathbf{{w}}^{*}=\lambda\,\boldsymbol{\Omega}_{0}\mathbf{{w}}^{*}, 
\label{eq:generalized_eigval}
\end{equation}
}
where the solution to (\ref{eq:generalized_eigval}) is
{\small
\begin{equation}
\mathbf{w}^{\star}\ = \sqrt{P_W} \cdot\mathbf{v}_{\max}\!\left(\boldsymbol{\Omega}_{0}^{-1}\boldsymbol{\Omega}_{1}\right),
\quad
R^{\star}=\lambda_{\max}\!\left(\boldsymbol{\Omega}_{0}^{-1}\boldsymbol{\Omega}_{1}\right),
\label{eq:optimal_w_soln}
\end{equation}}
where $\mathbf{v}_{\text{max}}(\boldsymbol{\Omega}_{0}^{-1}\boldsymbol{\Omega}_{1})$ is the eigenvector corresponding to the largest eigenvalue of $\boldsymbol{\Omega}_{0}^{-1}\boldsymbol{\Omega}_{1}$.  
\vspace{-0.2cm}
\section{Simulation Results}
\label{sec:simulations}

To demonstrate the validity of the collaboration design in (\ref{eq:optimal_w_soln}), for the detector in (\ref{eq:energy_detector}), we consider a simulation setup in which the IO waveform is modeled as an OFDM waveform with Quadrature-PSK alphabets. We consider a simulation example with $N=128$ and $D=32$. Each element of the complex subspace vector $\boldsymbol{\theta}_{\text{D}}$ is a Quadrature-PSK alphabet $[\boldsymbol{\theta}_{\text{D}}]_{l} \in \Big\{\pm \frac{1}{\sqrt{2}} \pm \frac{j}{\sqrt{2}} \Big\}$ for $l=1, \dots, D$. Without loss of generality, we assume that the $\boldsymbol{\Sigma}_{\text{D}} = \mathbf{I}_{\text{D}}$, so that the elements of $\boldsymbol{\theta}_{\text{D}}$ are all independent and identically distributed. The subspace $\boldsymbol{U}_{\text{D}}$ contains $D=32$ randomly selected columns of the discrete-Fourier transform (DFT) matrix, chosen from the $N=128$-dimensional DFT matrix. The noise in the receiver and surveillance channels is modeled as $\boldsymbol{\Sigma}_\text{r} = \boldsymbol{\Sigma}_\text{s} = \sigma_{\text{out}}^{2}(\mathbf{I}_{\text{N}}-\mathbf{U}_{\text{D}}\mathbf{U}^{\mathsf{H}}_{\text{D}}) + \sigma_{\text{in}}^{2}(\mathbf{U}_{\text{D}}\mathbf{U}^{\mathsf{H}}_{\text{D}})$, where the two terms correspond to the noise contributions from within the signal subspace and outside the signal subspace respectively. For the collaboration topology $\mathbf{A}$, we consider a network with $L = 8$ dual-channel receivers with $3$ neighbors for each transmitting receiver.


We use the plots in Fig.\ref{fig:collab_AF} to illustrate the benefits of inter-receiver collaboration over non-collaborative receivers, where the receivers only amplify their local CC statistic $c_i$, and transmit over the MAC channel to the FC. For the non-collaborative system (amplify-and-forward or AF), we assume that all $L=8$ receivers transmit to the FC, whereas for the collaborative systems, we assume that $M=5$ receivers transmit to the FC, after collaboration. The AF and collaborative systems are processed according to (\ref{eq:transform_subspace}) and (\ref{eq:cross_correl}). Also, as no collaboration takes place in the AF system, collaboration noise ($\epsilon_i$) is non-existent. We consider MAC fading coefficients $\mathbf{g}_{\text{AF}} = g_{\text{amp}} \cdot \mathbf{1}_L$, and $\mathbf{g}_{\text{collab}} = g_{\text{amp}} \cdot \mathbf{1}_M$, for the simulation experiments. The plots illustrate that the system employing collaboration does better than the AF system at lower values of $g_{\text{amp}}$, i.e., when the MAC channel coefficients are very low. When $g_{\text{amp}}$ is large, the $g_{\text{amp}}^{2}$ signal term dominates and MAC/collaboration noises are negligible, so AF and collaboration deliver essentially the same "energy" ($|z|^{2}$) to the FC, and yield nearly identical ROC curves. Also, the ROC curves are relatively more stable in collaborative systems across different $g_{\text{amp}}$ values, despite using fewer transmitters to the FC ($M = 5$).

\hspace{-0.4cm}Then, we illustrate the benefits of using the cross-correlation  defined in (\ref{eq:cross_correl}), which includes the subspace-transformed and whitened measurements. In Fig. \ref{fig:subspace_correl}, the receiver-operating characteristics (ROC) is plotted for the detection system that uses the subspace-aided cross-correlations in (\ref{eq:cross_correl}), and systems where the radar receivers only use noise whitening, before cross-correlating the measurements in the surveillance and reference channels as in \cite{liu2015performance, wei2021adaptive}. The detection performance improves when subspace transformations are applied to the channel measurements prior to CC. This is due to the fact that subspace transformations are effective in removing the noise-component from the receiver and surveillance noise. Also, the detection performance improves with larger target RCS variance $\sigma_{\alpha}^{2}$, as the surveillance channels, and the system employing subspace transformation is always more effective. 
\vspace{-0.3cm}
\section{Conclusion}
\label{sec:conclusions}
\vspace{-0.1cm}
In this work, we considered the target detection problem using a distributed passive-radar system where the illuminator-of-opportunity waveform subspace is known. This knowledge is used to improve the cross-correlation between reference and surveillance measurements. Receivers are allowed to collaborate with neighboring receivers and a subset of $M<L$ receivers transmit their measurements to the fusion center, after collaboration. A total of $2LN$ measurements collected across $L$ distributed receivers over $N$ sampling rounds are reduced to $M$ measurements after collaboration. These receivers transmit over a linear multiple-access channel channel to the fusion center (FC), and the received measurement is used to detect the target using a simple energy-detector test. Using the moments of the measurement at the FC, a surrogate performance metric was defined, and the collaboration weights are designed to maximize the detection performance. In future work, we extend this problem to scenarios where the illuminator waveform subspace is unknown.

\bibliographystyle{IEEEtran}
\bibliography{paper_refs}

\begin{thebibliography}{10}
\providecommand{\url}[1]{#1}
\csname url@samestyle\endcsname
\providecommand{\newblock}{\relax}
\providecommand{\bibinfo}[2]{#2}
\providecommand{\BIBentrySTDinterwordspacing}{\spaceskip=0pt\relax}
\providecommand{\BIBentryALTinterwordstretchfactor}{4}
\providecommand{\BIBentryALTinterwordspacing}{\spaceskip=\fontdimen2\font plus
\BIBentryALTinterwordstretchfactor\fontdimen3\font minus \fontdimen4\font\relax}
\providecommand{\BIBforeignlanguage}[2]{{%
\expandafter\ifx\csname l@#1\endcsname\relax
\typeout{** WARNING: IEEEtran.bst: No hyphenation pattern has been}%
\typeout{** loaded for the language `#1'. Using the pattern for}%
\typeout{** the default language instead.}%
\else
\language=\csname l@#1\endcsname
\fi
#2}}
\providecommand{\BIBdecl}{\relax}
\BIBdecl

\bibitem{berger2010signal}
C.~R. Berger, B.~Demissie, J.~Heckenbach, P.~Willett, and S.~Zhou, ``Signal processing for passive radar using ofdm waveforms,'' \emph{IEEE Journal of Selected Topics in Signal Processing}, vol.~4, no.~1, pp. 226--238, 2010.

\bibitem{polonen2013control}
K.~P{\"o}l{\"o}nen and V.~Koivunen, ``Control symbol based fluctuating target detection in dvb-t2 passive radar systems,'' in \emph{2013 IEEE Radar Conference (RadarCon13)}.\hskip 1em plus 0.5em minus 0.4em\relax IEEE, 2013, pp. 1--5.

\bibitem{gogineni2017passive}
S.~Gogineni, P.~Setlur, M.~Rangaswamy, and R.~R. Nadakuditi, ``Passive radar detection with noisy reference channel using principal subspace similarity,'' \emph{IEEE Transactions on Aerospace and Electronic Systems}, vol.~54, no.~1, pp. 18--36, 2017.

\bibitem{karthik2018improved}
A.~K. Karthik and R.~S. Blum, ``Improved detection performance for passive radars exploiting known communication signal form,'' \emph{IEEE Signal Processing Letters}, vol.~25, no.~11, pp. 1625--1629, 2018.

\bibitem{horstmann2020two}
S.~Horstmann, D.~Ram{\'\i}rez, and P.~J. Schreier, ``Two-channel passive detection of cyclostationary signals,'' \emph{IEEE Transactions on Signal Processing}, vol.~68, pp. 2340--2355, 2020.

\bibitem{mcwhorter2023passive}
L.~T. McWhorter, L.~Scharf, C.~Moore, and M.~Cheney, ``Passive multi-channel detection: A general first-order statistical theory,'' \emph{IEEE Open Journal of Signal Processing}, vol.~4, pp. 437--451, 2023.

\bibitem{ramirez2024passive}
D.~Ram{\'\i}rez, I.~Santamaria, and L.~L. Scharf, ``Passive detection with a multi-rank beamformer of a random signal common to two sensors,'' in \emph{2024 58th Asilomar Conference on Signals, Systems, and Computers}.\hskip 1em plus 0.5em minus 0.4em\relax IEEE, 2024, pp. 213--217.

\bibitem{wang2016canonical}
Y.~Wang, L.~L. Scharf, I.~Santamar{\'\i}a, and H.~Wang, ``Canonical correlations for target detection in a passive radar network,'' in \emph{2016 50th Asilomar Conference on Signals, Systems and Computers}.\hskip 1em plus 0.5em minus 0.4em\relax IEEE, 2016, pp. 1159--1163.

\bibitem{santamaria2017passive}
I.~Santamaria, L.~L. Scharf, J.~Via, H.~Wang, and Y.~Wang, ``Passive detection of correlated subspace signals in two mimo channels,'' \emph{IEEE Transactions on Signal Processing}, vol.~65, no.~20, pp. 5266--5280, 2017.

\bibitem{wang2017signal}
F.~Wang, H.~Li, X.~Zhang, and B.~Himed, ``Signal parameter estimation for passive bistatic radar with waveform correlation exploitation,'' \emph{IEEE Transactions on Aerospace and Electronic Systems}, vol.~54, no.~3, pp. 1135--1150, 2017.

\bibitem{liu2015performance}
J.~Liu, H.~Li, and B.~Himed, ``On the performance of the cross-correlation detector for passive radar applications,'' \emph{Signal Processing}, vol. 113, pp. 32--37, 2015.

\bibitem{wei2021adaptive}
J.~Wei, J.~Li, C.~Song, Z.~Xu, and K.~Ding, ``An adaptive fusion algorithm for multistatic and multichannel passive radar detection,'' in \emph{2021 IEEE Radar Conference (RadarConf21)}.\hskip 1em plus 0.5em minus 0.4em\relax IEEE, 2021, pp. 1--6.

\bibitem{ma2022compressive}
J.~Ma and J.~Jiang, ``Compressive subspace detectors based on sparse representation in multistatic passive radar systems,'' \emph{IEEE Transactions on Signal Processing}, vol.~70, pp. 5074--5086, 2022.

\bibitem{kar2013linear}
S.~Kar and P.~K. Varshney, ``Linear coherent estimation with spatial collaboration,'' \emph{IEEE Transactions on Information Theory}, vol.~59, no.~6, pp. 3532--3553, 2013.

\bibitem{zhang2018optimal}
S.~Zhang, S.~Liu, V.~Sharma, and P.~K. Varshney, ``Optimal sensor collaboration for parameter tracking using energy harvesting sensors,'' \emph{IEEE Transactions on Signal Processing}, vol.~66, no.~12, pp. 3339--3353, 2018.

\bibitem{cheng2021joint}
X.~Cheng, P.~Khanduri, B.~Chen, and P.~K. Varshney, ``Joint collaboration and compression design for distributed sequential estimation in a wireless sensor network,'' \emph{IEEE Transactions on Signal Processing}, vol.~69, pp. 5448--5462, 2021.

\end{thebibliography}

\end{document}